\documentclass{article}
\usepackage{arxiv}

\usepackage{cite}
\usepackage{amsmath,amssymb,amsfonts}
\usepackage{algorithm}
\usepackage{algorithmic}
\usepackage{graphicx}
\usepackage{textcomp}
\usepackage[table]{xcolor}
\usepackage{booktabs}
\usepackage{tabularx}
\usepackage{array}
\usepackage{ragged2e}
\usepackage{caption}
\usepackage{url}
\usepackage{tikz}
\usetikzlibrary{arrows.meta,positioning,shapes.geometric,calc,backgrounds,fit}
\usepackage{pgfplots}
\pgfplotsset{compat=1.18}
\usepackage[hidelinks]{hyperref}

\newcolumntype{L}{>{\RaggedRight\arraybackslash}X}

\def\BibTeX{{\rm B\kern-.05em{\sc i\kern-.025em b}\kern-.08em
    T\kern-.1667em\lower.7ex\hbox{E}\kern-.125emX}}

\definecolor{tableheader}{HTML}{1C355E}
\definecolor{rowtint}{HTML}{EEF2F8}
\definecolor{boxtint}{HTML}{E8EEF7}
\definecolor{gatetint}{HTML}{FBE9D6}

\newtheorem{definition}{Definition}
\newtheorem{proposition}{Design Proposition}

\newcommand{\spec}{\ensuremath{\mathcal{S}}}

\newcommand{\gen}{\ensuremath{G}}
\newcommand{\val}{\ensuremath{V}}
\newcommand{\accept}{\ensuremath{\mathcal{A}}}

\title{Reshaping the SDLC for Data- and AI-Centric Systems}

\author{
  Mamdouh Alenezi \\
  SDAIA Academy \\
  Saudi Data and Artificial Intelligence Authority (SDAIA) \\
  Riyadh, Saudi Arabia \\
  \texttt{malenezi@sdaia.gov.sa}
}

\date{}

\begin{document}
\maketitle

\begin{abstract}
The traditional Software Development Lifecycle (SDLC) rests on a deterministic premise: system behavior is fully determined by source code, and correctness can therefore be specified, implemented, and verified through code-centric practices. Data-intensive and AI-enabled systems break this premise, because their behavior is jointly determined by code, data, and learned models, and it degrades silently as the world drifts away from the data on which the models were trained. This paper examines how the integration of data engineering and software engineering practices---operationalized through DataOps, MLOps, and LLMOps---reshapes the SDLC for such systems. We make four contributions. First, we consolidate a fragmented literature spanning software engineering, data management, machine learning systems, and human-centered computing into a phase-structured account of lifecycle transformation, covering requirements, architecture, development, testing, deployment, monitoring, governance, and organization. Second, we give a lightweight formal treatment of the reshaped lifecycle: system behavior is defined over tripartite configurations of code, data, and model; requirements become evaluation-led specifications with probabilistic acceptance regions; and promotion is governed by statistically sound validation gates. Third, we synthesize these results into an adaptive, five-layer lifecycle framework---artifact, contract, gate, control, and governance layers---articulated as testable design propositions, in which maintenance becomes a closed-loop control problem over drifting configurations. Fourth, we propose a conceptual research model linking the degree of data-engineering integration to measurable lifecycle outcomes, and we critically appraise the evidence base, showing that while the direction of the transformation is well established, its magnitude remains insufficiently quantified. We close with a research agenda toward an empirically grounded, adaptive SDLC for data- and AI-centric systems.
\end{abstract}


\section{Introduction}
\label{sec:intro}

For half a century, the Software Development Lifecycle (SDLC) has provided the organizing scaffold of software engineering practice. From the waterfall model through iterative, spiral, and agile variants, the lifecycle has been premised on a single foundational assumption: that the behavior of a software system is fully determined by its source code. Requirements describe intended functionality; design decomposes that functionality into modules; development implements the modules; testing verifies that the code satisfies the requirements; deployment releases the code; and maintenance evolves the code. Every phase, every artifact, and every quality gate ultimately refers back to code as the singular source of truth. Continuous integration, delivery, and deployment compressed the interval between commit and release~\cite{shahin2017continuous}, but did not alter this artifact model: what flows through a conventional CI/CD pipeline is code and its compiled derivatives.

Data-intensive and AI-enabled systems break this assumption. In systems whose core behavior is learned from data---recommendation engines, fraud detection platforms, computer vision pipelines, conversational agents, and the rapidly proliferating class of large language model (LLM)-enabled and agentic applications---system behavior is jointly determined by \emph{code}, \emph{data}, and \emph{models}. The consequence, first articulated forcefully by Sculley et al., is that machine learning systems have a special capacity for incurring technical debt: they carry all the maintenance problems of traditional code plus an additional set of ML-specific issues arising from data dependencies, entanglement, and feedback loops~\cite{sculley2015hidden}. Amershi et al.'s landmark case study of software teams at Microsoft reached a complementary conclusion: data discovery, management, and versioning are fundamentally harder than code versioning; model building requires different skills than software construction; and AI components are entangled in ways that defeat modular decomposition~\cite{amershi2019software}. A decade of empirical software engineering research has replicated and extended these findings~\cite{martinez2022software,paleyes2022challenges,wan2021how,arpteg2018software,lwakatare2019taxonomy}, and the practical response of industry has been the systematic \emph{integration} of data engineering (DE) and software engineering (SE) practices---bidirectionally. Data-engineering disciplines (pipeline construction, data quality management, feature engineering, lineage) are embedded into the software process, and software-engineering disciplines (version control, code review, automated testing, continuous integration, infrastructure as code) are applied to data and model artifacts. This integration is operationalized under the umbrella paradigms DataOps~\cite{ereth2018dataops}, MLOps~\cite{kreuzberger2023mlops,john2021towards}, and, most recently, LLMOps, and it is part of a broader shift toward AI-native software engineering~\cite{he2025llm,abrahao2025software}.

This paper addresses the following research questions:

\begin{itemize}
  \item \textbf{RQ1 (Descriptive).} How does the integration of data engineering and software engineering practices reshape each phase of the SDLC for data-intensive and AI-enabled systems?
  \item \textbf{RQ2 (Outcome-oriented).} Through which constructs and mechanisms can the effect of this integration on lifecycle outcomes---efficiency, quality, reliability, and maintainability---be conceptualized and measured?
  \item \textbf{RQ3 (Design-oriented).} How can DE and SE practices be systematically integrated to establish an \emph{adaptive} SDLC, one whose structure responds to changing data, models, dependencies, and operational environments?
\end{itemize}

Our contributions are fourfold. \textbf{(C1)} We consolidate literature from software engineering venues (ICSE, TOSEM, TSE, EMSE, JSS), data management venues (VLDB, SIGMOD), machine learning systems venues (MLSys, NeurIPS), and human-centered computing venues (CHI, CSCW) into a single phase-structured account of the lifecycle transformation (Section~\ref{sec:phases}), answering RQ1. \textbf{(C2)} We provide a lightweight formalization of the reshaped lifecycle---tripartite configurations, evaluation-led specifications \spec{} with acceptance regions $\accept(\spec)$, generation operators \gen{}, and validation gates \val{}---that makes the conceptual transformations precise (Section~\ref{sec:model}). \textbf{(C3)} We synthesize the analysis into an adaptive five-layer lifecycle framework expressed as testable design propositions, framing maintenance as closed-loop control over drifting configurations (Section~\ref{sec:framework}), answering RQ3. \textbf{(C4)} We propose a conceptual research model with operationalizable constructs and critically appraise the evidence base, deriving a research agenda (Sections~\ref{sec:researchmodel} and~\ref{sec:agenda}), answering RQ2.

\section{Background and Related Paradigms}
\label{sec:background}

\subsection{The Deterministic Premise of the Traditional SDLC}

The traditional SDLC rests on three interlocking premises. The first is \emph{determinism}: given identical inputs and state, the system produces identical outputs, so correctness can be specified and verified. The second is \emph{code primacy}: the version-controlled repository is the single source of truth, and configuration management, build automation, and release engineering orbit around it. The third is \emph{phase-gated verification}: quality is assured by testing code against specifications, with unit, integration, system, and acceptance tests forming a well-understood pyramid~\cite{shahin2017continuous}.

Data-intensive systems---those whose primary engineering challenges concern the volume, velocity, variety, and quality of data---and AI-enabled systems---those in which at least one component's behavior is learned rather than programmed---violate all three premises. Such systems are \emph{probabilistic}: the mapping from input to output depends on learned parameters and, for generative models, on stochastic decoding, so correctness must be expressed as statistical acceptability within tolerance bounds~\cite{ozkaya2020what,giray2021software}. They are \emph{multi-artifact}: behavior depends on datasets, feature definitions, and model weights that live outside the code repository~\cite{sculley2015hidden,kreuzberger2023mlops}. And they are \emph{environmentally coupled}: because models encode a snapshot of the world's data distribution, concept drift and data drift silently degrade deployed behavior even when no artifact has changed~\cite{gama2014survey,lu2019learning}. A lifecycle for such systems must therefore be continuous by construction, not merely rapid.

\subsection{Data Engineering and Data-Centric AI}

Data engineering encompasses the design, construction, and operation of systems that collect, store, transform, and serve data at scale: ingestion pipelines, ETL/ELT workflows, warehouses and lakes, streaming platforms, feature computation, and the metadata systems that make these assets discoverable and trustworthy. Polyzotis et al.'s survey of data lifecycle challenges in production machine learning was an early bridge to software process research, cataloguing the data understanding, validation, cleaning, and enrichment tasks that dominate practitioner effort~\cite{polyzotis2018data}. Whang and Lee extended the analysis to deep learning, showing that data collection, labeling, and quality assurance constitute first-order engineering problems~\cite{whang2020data}. The data-centric AI movement, surveyed by Zha et al., reframes the object of iteration from the model to the dataset, systematizing training-data development, inference-data development, and data maintenance as engineering activities in their own right~\cite{zha2025data}. Sambasivan et al.'s study of ``data cascades'' in high-stakes AI supplied the human-centered corollary: undervalued data work compounds into downstream failures~\cite{sambasivan2021everyone}. Heck's mapping study of data engineering for AI systems confirms that organizations struggle precisely at the DE--SE boundary and identifies substantial lifecycle coverage gaps in the literature~\cite{heck2024data}.

\subsection{From DevOps to DataOps, MLOps, and LLMOps}

The integration of DE and SE has been operationalized through a family of ``Ops'' paradigms extending DevOps principles---automation, collaboration, measurement, flow---to data and model artifacts. DataOps applies agile and lean thinking to analytic data pipelines~\cite{ereth2018dataops}. MLOps extends the scope to model training, evaluation, deployment, and monitoring: Kreuzberger et al.\ derive its principles (CI/CD automation, workflow orchestration, reproducibility, tripartite versioning, continuous training, metadata tracking, feedback loops), components (pipelines, feature stores, model registries, monitoring), and roles~\cite{kreuzberger2023mlops}; John et al.\ contribute a maturity model~\cite{john2021towards}; Testi et al.\ a taxonomy and methodology~\cite{testi2022mlops}; Symeonidis et al.\ a tool-ecosystem mapping~\cite{symeonidis2022mlops}; and Tamburri a socio-technical framing of sustainable MLOps~\cite{tamburri2020sustainable}. LLMOps specializes the paradigm for systems built on large language models, where the unit of iteration is often a prompt, a retrieval corpus, or an agent orchestration graph: retrieval-augmented generation makes curated corpora and embedding pipelines part of the behavioral surface~\cite{lewis2020retrieval}; evaluation shifts toward benchmark suites, LLM-as-a-judge protocols~\cite{zheng2023judging,chang2024survey}, and hallucination measurement~\cite{ji2023survey}; and multi-agent orchestration introduces new lifecycle artifacts whose engineering is only beginning to be systematized~\cite{he2025llm}.

The economic rationale for integration is sharpest in the technical-debt literature: Sculley et al.\ enumerate debt categories peculiar to ML systems---entanglement, correction cascades, undeclared consumers, unstable data dependencies, glue code, pipeline jungles---and observe that only a small fraction of a production ML system is model code~\cite{sculley2015hidden}; Breck et al.\ operationalize the insight in the ML Test Score, a rubric of 28 tests spanning data, model, infrastructure, and monitoring~\cite{breck2017ml}. The cost of \emph{not} integrating manifests as unbounded, compounding maintenance cost~\cite{arpteg2018software,paleyes2022challenges,munappy2022data}.

\section{A Formal View of the Reshaped Lifecycle}
\label{sec:model}

We now make the conceptual transformations precise. The formalization is deliberately lightweight: its purpose is to expose \emph{why} each phase must change, and to give later design propositions unambiguous referents.

\begin{definition}[Tripartite configuration]
\label{def:config}
A \emph{system configuration} is a tuple
$c = (k, d, f, \theta, p, e)$,
where $k$ is a code commit, $d$ a data snapshot, $f$ a set of versioned feature (or transformation) definitions, $\theta$ a model artifact (weights or parameters), $p$ a prompt/orchestration artifact (possibly empty for non-LLM systems), and $e$ an execution environment. The observable behavior of the system is a function $B(c)$ of the whole configuration, not of $k$ alone.
\end{definition}

Definition~\ref{def:config} captures the first two core transformations: system state and logic are dictated by code, data, and model jointly, and the source of truth becomes multi-modal. Reproducing any historical behavior requires pinning the entire tuple, which is the raison d'\^etre of data version control, experiment tracking, model registries, and lineage systems~\cite{kreuzberger2023mlops,zaharia2018accelerating,barrak2021co,garcia2018context}.

\begin{definition}[Evaluation-led specification]
\label{def:spec}
An \emph{evaluation-led specification} is a tuple
$\spec = (D_{\mathrm{eval}}, \{m_j\}_{j=1}^{J}, \{\tau_j\}_{j=1}^{J})$,
where $D_{\mathrm{eval}}$ is a curated evaluation dataset (``golden set''), each $m_j$ is a measurable quality function (accuracy, calibration, latency, cost, fairness gap, groundedness, toxicity, \dots), and each $\tau_j$ is a threshold or tolerance band. The \emph{acceptance region} of \spec{} is
\[
\accept(\spec) \;=\; \bigl\{\, c \;:\; m_j\bigl(B(c), D_{\mathrm{eval}}\bigr) \trianglelefteq \tau_j \ \ \forall j \,\bigr\},
\]
where $\trianglelefteq$ denotes the appropriate ordering for each metric.
\end{definition}

Definition~\ref{def:spec} formalizes the shift from functional correctness to bounded probabilistic behavior: requirements are no longer logical postconditions but membership conditions in an acceptance region, expressed over data~\cite{habibullah2023non,studer2021towards}. The specification itself is an executable, versioned artifact---the AI-era analogue of acceptance-test-driven development.

\begin{definition}[Generation operator]
\label{def:gen}
A \emph{generation operator} $\gen : (d, f, h) \mapsto \theta$ maps a validated data snapshot $d$, feature definitions $f$, and a training configuration $h$ (hyperparameters, code, environment) to a model artifact $\theta$. In LLM-centric systems, \gen{} may instead (or additionally) produce a prompt/retrieval configuration $p$.
\end{definition}

\begin{definition}[Validation gate]
\label{def:gate}
A \emph{validation gate} is a decision procedure $\val_{\spec} : c \mapsto \{\textsf{accept}, \textsf{reject}\}$ estimating membership of $c$ in $\accept(\spec)$ from finite samples. A gate is \emph{statistically sound at level $\delta$} if $\Pr[\val_{\spec}(c) = \textsf{accept} \mid c \notin \accept(\spec)] \le \delta$. Renggli et al.\ show how such guarantees can be obtained with practical sample sizes inside a continuous-integration workflow~\cite{renggli2019continuous}.
\end{definition}

Because gates decide from samples, ``passing'' is a confidence-bounded judgment---the probabilistic analogue of a green build. Finally, drift enters the model as time dependence: the environment defines a data-generating distribution $P_t$ that evolves, so a configuration accepted against $\spec$ at time $t_0$ may exit the \emph{effective} acceptance region at $t_1 > t_0$ even though no artifact changed~\cite{gama2014survey,lu2019learning,rabanser2019failing}. The lifecycle objective is therefore an invariant to be \emph{maintained}, not a milestone to be reached:

\begin{equation}
\label{eq:invariant}
\forall t : \quad c_t \in \accept(\spec_t),
\end{equation}

where both the deployed configuration $c_t$ and the specification $\spec_t$ (whose evaluation data and thresholds must track the environment and the regulator) evolve. Equation~\eqref{eq:invariant} is the formal core of the ``adaptive SDLC'': Sections~\ref{sec:phases} and~\ref{sec:framework} describe the machinery---contracts, gates, continuous training, progressive delivery, monitoring---by which practice attempts to maintain it.

\begin{figure}[t]
\centering
\resizebox{\textwidth}{!}{%
\begin{tikzpicture}[
  font=\small,
  loop/.style={rectangle, rounded corners=3pt, draw=tableheader, thick, fill=boxtint,
               minimum width=3.4cm, minimum height=1.9cm, align=center},
  iface/.style={rectangle, rounded corners=2pt, draw=tableheader!70, fill=gatetint,
                minimum width=2.5cm, minimum height=1.15cm, align=center, font=\footnotesize},
  base/.style={rectangle, rounded corners=3pt, draw=tableheader, thick, fill=rowtint,
               minimum height=0.85cm, align=center, font=\footnotesize},
  arr/.style={-{Stealth[length=2.6mm]}, thick, tableheader},
  feed/.style={-{Stealth[length=2.6mm]}, thick, tableheader!70, dashed}
]
\node[loop] (data)  {\textbf{Data Loop}\\[1pt]\footnotesize ingestion $\cdot$ transformation\\[-1pt]\footnotesize validation $\cdot$ curation};
\node[iface, right=0.55cm of data] (c1) {Data contracts\\Feature store};
\node[loop, right=0.55cm of c1] (model) {\textbf{Model Loop}\\[1pt]\footnotesize experimentation $\cdot$ training \gen\\[-1pt]\footnotesize evaluation \val $\cdot$ registration};
\node[iface, right=0.55cm of model] (c2) {Model registry\\Serving API};
\node[loop, right=0.55cm of c2] (code) {\textbf{Code Loop}\\[1pt]\footnotesize sprints $\cdot$ review $\cdot$ CI/CD\\[-1pt]\footnotesize progressive delivery};
\node[base, below=0.75cm of c1.south -| model.south, minimum width=14.6cm] (meta)
  {Shared metadata, lineage, and observability plane (every artifact of $c=(k,d,f,\theta,p,e)$ traceable end to end)};
\draw[arr] (data) -- (c1);
\draw[arr] (c1) -- (model);
\draw[arr] (model) -- (c2);
\draw[arr] (c2) -- (code);
\draw[arr, tableheader!60] (data.south) -- (data.south |- meta.north);
\draw[arr, tableheader!60] (model.south) -- (model.south |- meta.north);
\draw[arr, tableheader!60] (code.south) -- (code.south |- meta.north);
\draw[feed] (code.north) .. controls +(0,1.1) and +(0,1.1) .. node[above, font=\footnotesize, text=tableheader]{production telemetry, user feedback, outcome labels} (data.north);
\end{tikzpicture}}
\caption{The reshaped SDLC as a multi-loop ecosystem. Three semi-independent iteration loops with irreducibly different tempos are coupled through explicit contract interfaces rather than merged into a single cadence; a shared metadata, lineage, and observability plane underpins reproducibility and audit; and instrumented feedback from production closes the outer loop~\cite{kreuzberger2023mlops,garcia2018context,orr2021managing}.}
\label{fig:loops}
\end{figure}
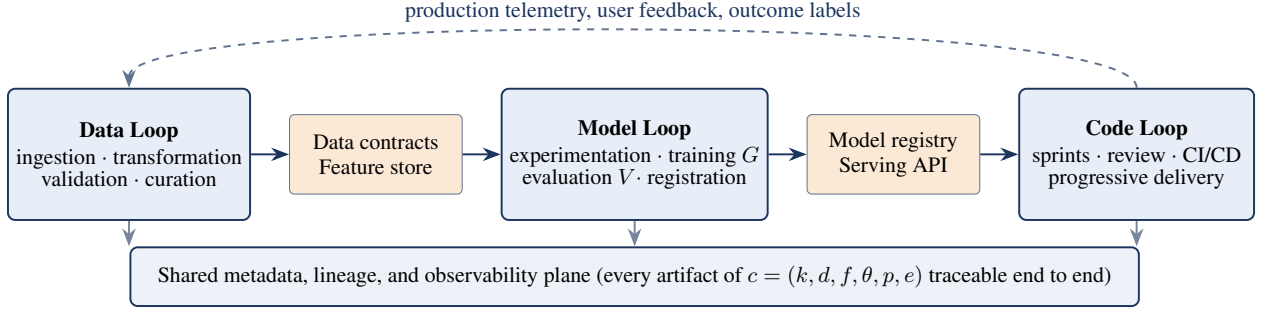

\subsection{Dual-Track and Triple-Track Iteration}

Operationally, the integrated lifecycle runs at least two, and in AI-enabled systems three, semi-independent loops (Figure~\ref{fig:loops}). The \emph{code loop} is governed by conventional agile practice: sprints, feature flags, semantic versioning, review, CI. The \emph{data loop} follows asynchronous rhythms of ingestion, transformation, validation, feature engineering, and curation, orchestrated by workflow engines and validated by data quality frameworks; its cadence is set by upstream availability and freshness requirements rather than sprint boundaries~\cite{polyzotis2018data,whang2020data}. The \emph{model loop} iterates over experimentation, training (\gen), evaluation (\val), and registration, with cadence set by drift signals, evaluation regressions, and business events~\cite{kreuzberger2023mlops,gama2014survey}. Integration is achieved not by merging the loops---their tempos differ irreducibly---but by coupling them through explicit interfaces: data contracts and feature stores between the data and model loops, registries and serving APIs between the model and code loops, and shared metadata, lineage, and observability infrastructure across all three~\cite{orr2021managing,zaharia2018accelerating,garcia2018context}. Process models formalize the resulting topology: CRISP-ML(Q) attaches quality assurance to every phase from business and data understanding through monitoring~\cite{studer2021towards}; the MLOps end-to-end architecture renders it as automated workflow~\cite{kreuzberger2023mlops}; and ISO/IEC~5338 embeds AI-specific processes into the international lifecycle standards family~\cite{iso5338}. In all of them, ``maintenance'' is not a terminal phase but the permanent operating mode.

\section{Phase-by-Phase Transformation of the SDLC}
\label{sec:phases}

Table~\ref{tab:shifts} summarizes the five conceptual transformations that underlie every phase-level change; the subsections that follow trace their consequences through the lifecycle.

\begin{table}[t]
\centering
\caption{Core conceptual transformations of the SDLC under DE--SE integration.}
\label{tab:shifts}
\small
\begin{tabularx}{\textwidth}{@{}p{2.5cm} L L L@{}}
\toprule
\rowcolor{tableheader}
\textcolor{white}{\textbf{Dimension}} &
\textcolor{white}{\textbf{Traditional SDLC}} &
\textcolor{white}{\textbf{Data/AI-centric SDLC}} &
\textcolor{white}{\textbf{Operational shift}} \\
\midrule
System state and logic & Deterministic; code dictates state & Probabilistic; $B(c)$ depends on code, data, and model jointly (Def.~\ref{def:config}) & From functional correctness to bounded probabilistic behavior~\cite{ozkaya2020what,giray2021software} \\
\rowcolor{rowtint}
Source of truth & Git repository (code versioning) & Code + data snapshots + features + weights + prompts, linked by lineage & Multi-modal, coupled versioning~\cite{sculley2015hidden,kreuzberger2023mlops,barrak2021co} \\
Testing paradigm & Unit, integration, end-to-end tests & Schema validation, drift detection, statistical evaluation, behavioral tests, evals (Def.~\ref{def:gate}) & Continuous validation of data and model quality alongside code~\cite{breck2017ml,zhang2022machine} \\
\rowcolor{rowtint}
CI/CD pipeline & Build $\rightarrow$ test $\rightarrow$ deploy & CI + continuous training + continuous evaluation + progressive delivery (CI/CD/CT/CM) & Automated retraining loops and evaluation-gated promotion~\cite{kreuzberger2023mlops,renggli2019continuous} \\
Monitoring and feedback & System health (CPU, latency, errors) & + data drift, concept drift, prediction quality, fairness, hallucination, cost & Real-time observability of data decay and output distributions~\cite{gama2014survey,schroder2022monitoring} \\
\bottomrule
\end{tabularx}
\end{table}

\subsection{Requirements Engineering and Planning}
\label{sec:req}

Requirements engineering is the phase least studied and among the most disrupted. Vogelsang and Borg's interview study found that requirements for ML systems must cover new object types---training data, runtime data, and model behavior---and demand new analyst competencies, including reasoning about data availability, label quality, and achievable performance before commitments are made~\cite{vogelsang2019requirements}. Villamizar et al.'s systematic mapping confirms that conventional elicitation, specification, and validation techniques transfer poorly and that ML-specific proposals remain weakly evaluated~\cite{villamizar2021requirements}.

In the integrated lifecycle the requirements baseline expands along four axes. \emph{Data requirements} specify sources, access rights, volume, velocity, schema, semantics, freshness, retention, and quality thresholds, each a testable obligation on the pipeline. \emph{Model success criteria} define measurable targets for accuracy, calibration, latency, cost, robustness, fairness, and explainability, expressed as the thresholds $\tau_j$ of Definition~\ref{def:spec} rather than binary conditions~\cite{studer2021towards,habibullah2023non}. \emph{Data contracts} formalize producer--consumer agreements about schema, semantics, service levels, and quality before pipeline code is written~\cite{dehghani2022data,machado2022data}. \emph{Compliance requirements} encode privacy, provenance, and regulatory constraints on both data handling and model outputs~\cite{nist2023ai,euaiact2024}. Habibullah et al.\ show empirically that classic non-functional-requirement frameworks are perceived as inadequate: qualities such as fairness and retrainability lack agreed definitions and measurement scales, and NFRs for ML must often be defined \emph{over data} rather than over code~\cite{habibullah2023non}. For LLM-enabled systems the phase increasingly takes the form of \emph{evaluation-led development}: requirements are encoded directly as the executable specification \spec---golden datasets, scoring rubrics, threshold policies---versioned and executed throughout the system's life~\cite{chang2024survey,zheng2023judging}. Planning changes accordingly: feasibility analysis includes data audits and baseline experiments, and estimation budgets for acquisition, labeling, and cleaning, which routinely dominate effort~\cite{polyzotis2018data,sambasivan2021everyone}.

\subsection{Architecture and System Design}
\label{sec:arch}

The dominant design stance shifts from \emph{schema-first} to \emph{contract-first}. Data mesh generalizes the stance into an architectural paradigm---domain-oriented ownership, data as a product with SLAs and quality guarantees, self-serve infrastructure, federated computational governance---so that the architectural unit becomes the versioned, documented, discoverable \emph{data product} with the interface rigor of a software API~\cite{dehghani2022data,machado2022data}.

The integrated lifecycle has produced a canonical component set absent from conventional systems. \emph{Feature stores} provide versioned feature definitions with dual serving paths---batch for training, low-latency for inference---eliminating training--serving skew by construction~\cite{kreuzberger2023mlops,orr2021managing}. \emph{Model registries} record versions, lineage, evaluation results, and deployment stage~\cite{zaharia2018accelerating}. \emph{Pipeline orchestrators} express data and training workflows as versioned, testable DAG code, replacing the ``pipeline jungles'' identified as a principal debt source~\cite{sculley2015hidden,kreuzberger2023mlops}. \emph{Metadata, catalog, and lineage systems} record the provenance graph connecting datasets, transformations, features, runs, and models, so that any prediction can be audited back to the exact commit, pipeline run, and data snapshot that produced it~\cite{garcia2018context,herschel2017survey}. Production platforms such as TFX demonstrate the integrated pattern at scale, with data validation and model analysis as first-class pipeline stages~\cite{baylor2017tfx,breck2019data}.

Two design principles recur. First, \emph{decoupling of serving and training layers}: real-time inference is isolated from asynchronous batch computation and retraining via event-driven messaging and service boundaries, so heavy data-plane workloads cannot degrade the request path and models can be swapped without redeploying applications~\cite{paleyes2022challenges,kreuzberger2023mlops}. Second, \emph{deterministic fallback mechanisms}: rule-based heuristics, cached responses, or simpler baseline models act as governed degradation paths when model outputs exceed confidence boundaries or fail evaluation constraints~\cite{breck2017ml,paleyes2022challenges}. LLM-enabled systems add AI-native elements---retrieval-augmented generation coupling models to vector stores and embedding pipelines~\cite{lewis2020retrieval}, agent orchestration graphs and tool contracts~\cite{he2025llm}, and prompt-management layers promoted with code-like discipline. Model-driven engineering is beginning to respond with DSLs and modeling support, though no widely adopted method yet spans the whole lifecycle~\cite{radler2024bridging}. Finally, reproducibility becomes a \emph{designed} property: immutable snapshots, containerized environments, and complete metadata capture guarantee that any historical configuration $c$ can be reconstructed~\cite{kreuzberger2023mlops,garcia2018context}, while backward-compatible contract evolution counters the unstable-dependency and undeclared-consumer debt patterns~\cite{sculley2015hidden}.

\subsection{Development and Implementation}
\label{sec:dev}

The most visible marker of integration is the extension of the ``as code'' discipline to every artifact: pipelines, transformations, feature definitions, infrastructure, training workflows, evaluation suites, prompts, and orchestration graphs are version-controlled, reviewable, and testable, with infrastructure-as-code as the enabling substrate~\cite{rahman2019systematic}. The pull request becomes the universal unit of change for the tripartite lifecycle. Serban et al.'s survey of 29 engineering best practices grounds the convergence empirically: adoption correlates with practitioners' ability to achieve agility, quality, and traceability, and traditional SE practices rank among the most effective even in ML contexts~\cite{serban2020adoption}; Wan et al.\ document how engineers actively import SE discipline to tame ML-induced changes across the lifecycle~\cite{wan2021how}.

Reproducible development requires \emph{coupled} versioning of code, data, and model. Data version control brings Git-like semantics to datasets and model files; Barrak et al.'s study of DVC projects shows data-versioning artifacts co-evolving with source code and tests, with real but manageable overhead~\cite{barrak2021co}. Experiment tracking records, for every run, the code version, data version, hyperparameters, environment, and metrics; the registry links accepted experiments to deployable artifacts with evidence-governed stage transitions~\cite{zaharia2018accelerating,garcia2018context}. A persistent tension is the \emph{notebook-to-production gap}: Pimentel et al.'s analysis of over one million Jupyter notebooks demonstrates that most are non-reproducible~\cite{pimentel2019large}; the integrated lifecycle responds with an explicit promotion path from exploratory notebooks into modular, tested pipeline code~\cite{studer2021towards,serban2020adoption}.

Data engineering contributes its most distinctive development practice: \emph{shift-left data quality}. Declarative expectations---schema conformity, nulls, ranges, uniqueness, referential integrity, distributional stability---are written alongside pipeline code and executed at ingestion, intercepting malformed data before feature computation, training, or inference~\cite{schelter2018automating,breck2019data}. The targeted failure mode, \emph{silent data corruption}, in which upstream changes break downstream inference without raising execution errors, is among the most frequently reported production incidents~\cite{paleyes2022challenges,polyzotis2018data}. The complementary practice is telemetry design: instrumentation is built into runtime systems from the outset so that production interactions are captured as well-schematized data feeds for future training and evaluation~\cite{kreuzberger2023mlops,breck2017ml}.

\subsection{Testing and Quality Assurance}
\label{sec:test}

Traditional testing verifies code against specifications; for learned components there is no complete specification, expected outputs are unknown for most inputs (the oracle problem), and behavior depends on training data that testing never touches. Zhang et al.'s survey maps the resulting field by testing properties, components under test, and workflow~\cite{zhang2022machine}; Riccio et al.\ catalogue the empirical maturity of proposed techniques~\cite{riccio2020testing}. Both converge on the multi-layered stack the integrated lifecycle institutionalizes: \emph{data testing} (schema, completeness, distributions)~\cite{breck2019data,schelter2018automating}; \emph{pipeline testing} (transformation logic, idempotency, failure recovery)~\cite{kreuzberger2023mlops}; \emph{model evaluation} (held-out performance, calibration, robustness, slice-level behavior, fairness)~\cite{breck2017ml,mehrabi2021survey}; \emph{behavioral testing}, where CheckList-style minimum-functionality, invariance, and directional tests expose systematic failures invisible to aggregate metrics~\cite{ribeiro2020beyond}; \emph{metamorphic testing}, asserting relations between outputs under input transformations to circumvent the oracle problem~\cite{segura2016survey}; and, for LLM systems, \emph{evaluation suites} integrating golden datasets, rubric scoring, hallucination detection, and LLM-as-a-judge protocols---whose own biases must in turn be validated~\cite{zheng2023judging,ji2023survey,chang2024survey}.

The decisive structural change is that these layers are wired into CI as automated gates (Definition~\ref{def:gate}): promotion is blocked when schema contracts are violated, drift is detected in training data, holdout evaluation drops below threshold, or fairness and robustness constraints regress~\cite{breck2017ml,breck2019data,kreuzberger2023mlops}. Renggli et al.\ supply the statistical foundations, showing how model-quality conditions can be tested inside CI with rigorous guarantees at practical sample sizes~\cite{renggli2019continuous}, and the ML Test Score provides a maturity yardstick across 28 tests of data, model, infrastructure, and monitoring~\cite{breck2017ml}. Quality assurance is thereby redefined: a system can be correct in code yet fail QA because its data violates expectations or its model has degraded---and the pipeline detects all three conditions continuously.

\subsection{Deployment and Release Engineering}
\label{sec:deploy}

Deployment releases three coupled artifact classes---application code, data pipelines, and models---whose versions must remain mutually consistent. The pipeline extends classical CI/CD~\cite{shahin2017continuous} with \emph{continuous training}: when monitoring detects drift or scheduled triggers fire, orchestrated pipelines fetch fresh data, validate it, retrain candidates via \gen, evaluate them against \val, and register or deploy them, with human approval reserved for consequential cases~\cite{kreuzberger2023mlops,john2021towards}. Because offline evaluation cannot fully predict online behavior, models and prompts are released progressively: \emph{shadow deployments} log a candidate's outputs on live traffic without serving them; \emph{canary releases} shift a small traffic fraction and compare quality, latency, and cost; and \emph{A/B tests} measure business-level effects~\cite{paleyes2022challenges,kreuzberger2023mlops}. Shankar et al.'s interview study captures the practitioner rationale in its title---``we have no idea how models will behave in production until production''---and documents staged deployment, rapid rollback, and continual online evaluation as institutionalized compensations for the epistemic limits of offline testing~\cite{shankar2024we}.

Two mechanisms are distinctive. \emph{Training--serving skew elimination}: feature stores and shared transformation code guarantee that the exact transformations applied during training are applied at inference, removing a classic silent-divergence failure architecturally rather than testing it away~\cite{orr2021managing,baylor2017tfx}. \emph{Pipeline-aware rollback}: rolling back a model may require rolling back feature definitions, embedding indices, or transformations on which it depends, so the registry and lineage system must record the full dependency closure of every deployed configuration and treat the tuple $c$---not the model alone---as the unit of rollback~\cite{kreuzberger2023mlops,herschel2017survey,paleyes2022challenges}.

\subsection{Monitoring, Operations, and Maintenance}
\label{sec:monitor}

Operations expand from application performance monitoring to comprehensive \emph{data and AI observability}: freshness, volume, schema, and distributional properties of every feeding pipeline; prediction distributions, confidence, slice-level quality, and fairness of every model; and, for LLM and agent systems, token usage, cost, latency percentiles, trace histories, guardrail triggers, and user feedback~\cite{schroder2022monitoring,polyzotis2018data,ji2023survey}. The scientific core of maintenance is \emph{drift}: Gama et al.\ established the taxonomy and adaptive-learning architectures~\cite{gama2014survey}, Lu et al.\ systematized detection and adaptation methods~\cite{lu2019learning}, and Rabanser et al.\ showed empirically that many practical shift-detection methods fail quietly, implying that drift detection is itself an engineered, tested component~\cite{rabanser2019failing}.

In lifecycle terms, drift converts maintenance from a reactive, ticket-driven activity into the closed-loop control problem of Equation~\eqref{eq:invariant}: monitors estimate the deployed configuration's distance from its validated operating envelope; controllers decide among continue, retrain, rollback, fallback, or escalate; and the continuous-training pipelines of Section~\ref{sec:deploy} act as actuators~\cite{kreuzberger2023mlops,gama2014survey}. Sculley et al.'s warning about feedback loops applies with full force: systems that influence the data they later train on can amplify their own biases, so the feedback pipelines that make systems self-improving must themselves be governed, sampled, and audited~\cite{sculley2015hidden,breck2017ml}. Operational success in this regime depends more on the maturity of pipelines, automation, observability, and governance than on model accuracy alone~\cite{phan2025operationalizing}; and maintenance planning budgets for pipeline evolution, schema migration, corpus refresh, re-labeling, retraining, and evaluation-suite upkeep---recurring costs that case studies confirm dominate total cost of ownership~\cite{munappy2022data,paleyes2022challenges}.

\subsection{Governance, Security, and Responsible AI}
\label{sec:gov}

Governance is re-implemented \emph{as} engineering: policies become executable checks, documentation becomes generated artifacts, and auditability becomes a property guaranteed by lineage infrastructure. Datasheets for datasets standardize documentation of motivation, composition, collection, preprocessing, uses, and limitations~\cite{gebru2021datasheets}; model cards document intended use, evaluation conditions, slice-level performance, and ethical considerations~\cite{mitchell2019model}; Hutchinson et al.\ extend both into a lifecycle accountability framework producing audit-ready artifacts at every stage of dataset development~\cite{hutchinson2021towards}. Provenance supplies the enforcement substrate~\cite{herschel2017survey}, fairness monitoring runs continuously in production rather than once at release~\cite{mehrabi2021survey}, and the expanded attack surface---training-data poisoning, model theft and evasion, prompt injection and retrieval poisoning---brings security engineering inside the pipeline~\cite{martinez2022software,zhang2022machine,ji2023survey}. External instruments increasingly mandate what the integrated lifecycle already enables: the EU AI Act's obligations on data governance, documentation, logging, human oversight, robustness, and post-market monitoring map directly onto the mechanisms of Sections~\ref{sec:req}--\ref{sec:monitor}~\cite{euaiact2024}; the NIST AI Risk Management Framework presupposes continuous measurement and feedback~\cite{nist2023ai}; and ISO/IEC~5338 formalizes AI-specific lifecycle processes~\cite{iso5338}. Regulatory compliance and engineering excellence have thus converged on the same infrastructure.

\subsection{Organizational Structures, Roles, and Culture}
\label{sec:org}

The transformation is as much organizational as technical. Kim et al.\ documented the initial condition---data specialists embedded in engineering organizations with distinct tools and quality norms~\cite{kim2018data}---and Nahar et al.\ identified the collaboration points where friction concentrates: requirements negotiation, data handoffs, and integration handoffs, with unclear responsibilities at these points a dominant failure source~\cite{nahar2022collaboration}. The silos between data scientist, data engineer, and software engineer dissolve into cross-functional product teams supported by platform teams, with hybrid roles---ML engineer, AI engineer, full-stack data engineer---institutionalizing the integration~\cite{amershi2019software,kreuzberger2023mlops,phan2025operationalizing}. Teams shift from project delivery to product stewardship of learning systems, supported by DataOps/MLOps culture~\cite{ereth2018dataops,john2021towards}. Abrah\~ao et al.\ argue the transformation must remain human-centered, redesigned around human judgment, ethics, and human--AI collaboration rather than automation for its own sake~\cite{abrahao2025software}; and the educational corollary---curricula joining SE rigor, DE fluency, and ML literacy---is poorly served by teaching the disciplines in isolation~\cite{kastner2020teaching}. Sambasivan et al.'s data-cascades finding carries the cultural warning: as long as data work is treated as low-status labor rather than engineering, the integrated lifecycle is starved of the practices on which everything downstream depends~\cite{sambasivan2021everyone}.

\section{An Adaptive, Integrated Lifecycle Framework}
\label{sec:framework}

\subsection{Maintenance as Closed-Loop Control}

\begin{figure}[t]
\centering
\begin{tikzpicture}[
  font=\small,
  node distance=0.9cm and 1.0cm,
  blk/.style={rectangle, rounded corners=3pt, draw=tableheader, thick, fill=boxtint,
              minimum width=2.9cm, minimum height=1.25cm, align=center, font=\footnotesize},
  gate/.style={rectangle, rounded corners=3pt, draw=tableheader, thick, fill=gatetint,
              minimum width=2.9cm, minimum height=1.25cm, align=center, font=\footnotesize},
  arr/.style={-{Stealth[length=2.6mm]}, thick, tableheader}
]
\node[blk] (plant) {\textbf{Plant}\\deployed configuration $c_t$\\(system + pipelines + models)};
\node[blk, right=of plant] (sense) {\textbf{Sensors}\\data \& AI observability\\drift signals, quality, cost};
\node[gate, right=of sense] (comp) {\textbf{Comparators}\\gates $\val_{\spec_t}$, drift detectors\\$c_t \in \accept(\spec_t)$?};
\node[blk, below=of comp] (ctrl) {\textbf{Controller}\\policy: continue / retrain /\\rollback / fallback / escalate};
\node[blk, below=of plant] (act) {\textbf{Actuators}\\CT pipelines \gen, progressive\\delivery, fallback, human review};
\node[above=0.45cm of comp, font=\footnotesize, text=tableheader] (spec) {specification $\spec_t$ (evals, thresholds, regulation)};
\draw[arr] (plant) -- (sense);
\draw[arr] (sense) -- (comp);
\draw[arr] (spec) -- (comp);
\draw[arr] (comp) -- (ctrl);
\draw[arr] (ctrl) -- (act);
\draw[arr] (act) -- (plant);
\end{tikzpicture}
\caption{The adaptive lifecycle as a closed control loop maintaining the invariant $c_t \in \accept(\spec_t)$ of Equation~\eqref{eq:invariant}. Maintenance ceases to be a terminal phase and becomes the permanent operating mode of the system.}
\label{fig:control}
\end{figure}
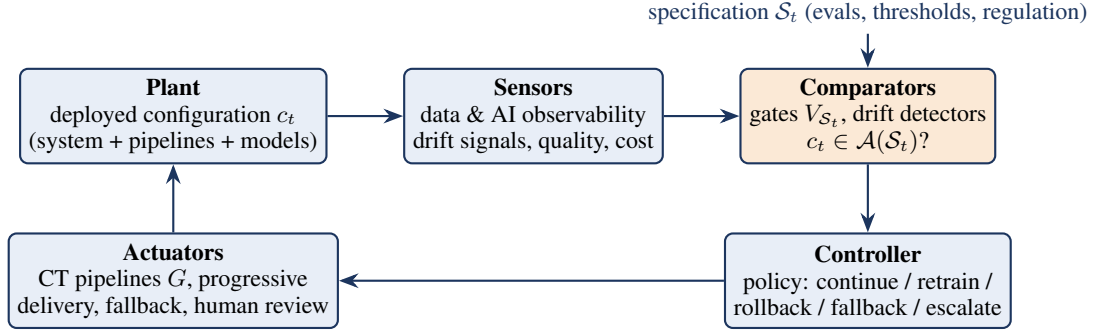

The phase analysis converges on a claim stronger than ``the SDLC has more steps'': AI-enabled systems require a lifecycle whose \emph{structure} responds to change without human re-planning of the process itself. The appropriate frame is control-theoretic (Figure~\ref{fig:control}): the deployed system, its pipelines, and its models constitute a plant; observability constitutes the sensor layer; evaluation gates and drift detectors constitute comparators against the current specification $\spec_t$; and continuous training, progressive delivery, fallback, rollback, and human escalation constitute actuators. An \emph{adaptive SDLC} is one in which this loop is explicit, engineered, tested, and governed rather than emergent and ad hoc~\cite{kreuzberger2023mlops,gama2014survey,studer2021towards,phan2025operationalizing}.

\subsection{A Five-Layer Reference Structure}

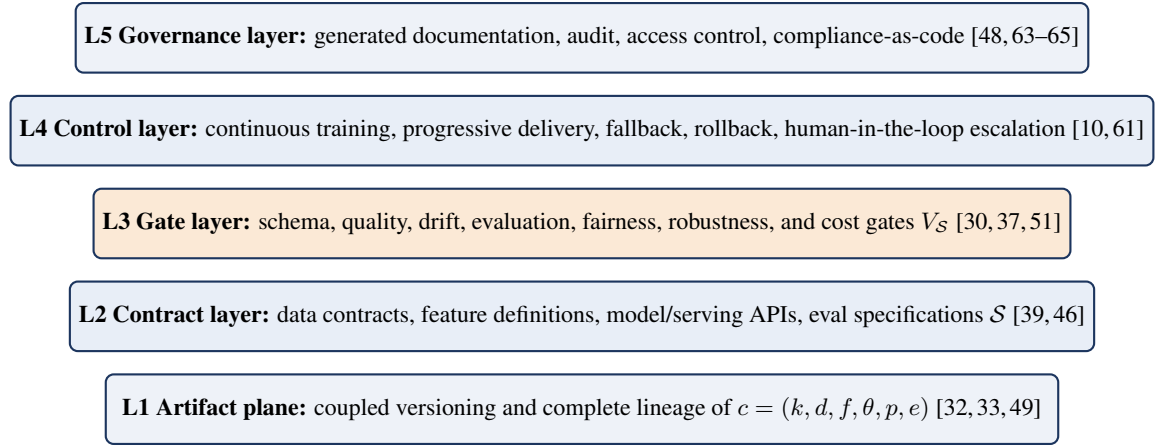
\begin{figure}[t]
\centering
\begin{tikzpicture}[
  font=\small,
  layer/.style={rectangle, rounded corners=3pt, draw=tableheader, thick,
                minimum width=12.6cm, minimum height=0.92cm, align=center, font=\footnotesize}
]
\node[layer, fill=rowtint]  (l5) {\textbf{L5 Governance layer:} generated documentation, audit, access control, compliance-as-code~\cite{gebru2021datasheets,mitchell2019model,hutchinson2021towards,euaiact2024}};
\node[layer, fill=boxtint,  below=0.28cm of l5] (l4) {\textbf{L4 Control layer:} continuous training, progressive delivery, fallback, rollback, human-in-the-loop escalation~\cite{kreuzberger2023mlops,shankar2024we}};
\node[layer, fill=gatetint, below=0.28cm of l4] (l3) {\textbf{L3 Gate layer:} schema, quality, drift, evaluation, fairness, robustness, and cost gates $\val_{\spec}$~\cite{breck2017ml,breck2019data,renggli2019continuous}};
\node[layer, fill=boxtint,  below=0.28cm of l3] (l2) {\textbf{L2 Contract layer:} data contracts, feature definitions, model/serving APIs, eval specifications \spec~\cite{machado2022data,orr2021managing}};
\node[layer, fill=rowtint,  below=0.28cm of l2] (l1) {\textbf{L1 Artifact plane:} coupled versioning and complete lineage of $c=(k,d,f,\theta,p,e)$~\cite{zaharia2018accelerating,barrak2021co,herschel2017survey}};
\end{tikzpicture}
\caption{Five-layer reference structure of the adaptive integrated lifecycle.}
\label{fig:layers}
\end{figure}

Synthesizing Sections~\ref{sec:model} and~\ref{sec:phases}, the adaptive integrated lifecycle comprises five layers (Figure~\ref{fig:layers}), which we articulate as testable design propositions.

\begin{proposition}[Tripartite artifact plane]
\label{prop:artifact}
Every deployable behavior shall be reproducible from a fully versioned configuration $c=(k,d,f,\theta,p,e)$ with complete lineage; any artifact not under coupled version control is outside the engineering process and constitutes latent debt~\cite{sculley2015hidden,kreuzberger2023mlops,garcia2018context}.
\end{proposition}

\begin{proposition}[Contract-mediated decoupling]
\label{prop:contract}
The data, model, and code loops shall be coupled exclusively through explicit, versioned contracts (data contracts, feature definitions, model/serving APIs, eval specifications), so that each loop can iterate at its own tempo without cascading breakage~\cite{machado2022data,orr2021managing,nahar2022collaboration}.
\end{proposition}

\begin{proposition}[Uniform statistical gating]
\label{prop:gate}
Promotion of any configuration change---code, data, feature, model, or prompt---shall pass validation gates $\val_{\spec}$ that are statistically sound at a declared level $\delta$ and applied uniformly at integration, promotion, and runtime~\cite{renggli2019continuous,breck2017ml,breck2019data}.
\end{proposition}

\begin{proposition}[Engineered adaptation]
\label{prop:control}
Responses to monitored signals---retraining, rollback, fallback, escalation---shall be implemented as governed, tested control actions maintaining $c_t \in \accept(\spec_t)$, with human approval bound to consequence rather than to routine~\cite{gama2014survey,kreuzberger2023mlops,shankar2024we}.
\end{proposition}

\begin{proposition}[Governance by construction]
\label{prop:gov}
Documentation, audit trails, access control, and compliance evidence shall be generated and enforced by the same infrastructure that delivers the system, so that regulatory obligations are satisfied as a by-product of engineering practice~\cite{hutchinson2021towards,herschel2017survey,euaiact2024,nist2023ai}.
\end{proposition}

\section{A Conceptual Research Model}
\label{sec:researchmodel}

For empirical investigation, the transformation can be conceptualized as a causal chain:
\begin{center}
\emph{Integration of DE practices} $\rightarrow$ \emph{Integrated engineering practices (data + code + ML + DevOps + governance)} $\rightarrow$ \emph{Reshaped SDLC (adaptive, multi-loop, evaluation-gated)} $\rightarrow$ \emph{Software and system outcomes.}
\end{center}
The independent construct---degree of DE-practice integration---can be operationalized through practice-adoption instruments~\cite{serban2020adoption}, the ML Test Score~\cite{breck2017ml}, and MLOps maturity stages~\cite{john2021towards}, covering data versioning and lineage, data quality engineering, automated pipelines, data/ML testing, data/ML CI/CD, infrastructure as code, continuous monitoring, drift management, reproducibility, and data/AI governance. The dependent constructs---development efficiency, software and data quality, system reliability, maintainability, deployment frequency, time to recovery, pipeline reliability, and model lifecycle efficiency---can draw on established software-delivery performance measures (deployment frequency, lead time, change-failure rate, time to restore)~\cite{forsgren2018accelerate} extended with data-plane analogues: pipeline SLA adherence, data-incident rate, time-to-detect drift, and retraining lead time. Plausible moderators include system criticality, regulatory exposure, organizational maturity, and team topology; plausible mediators include reproducibility, observability coverage, and collaboration quality~\cite{john2021towards,serban2020adoption,nahar2022collaboration}. The model supports both variance studies (does more integration predict better outcomes?) and process studies (through which mechanisms?), giving RQ2 measurable form.

\section{Evidence Base, Limitations, and Research Agenda}
\label{sec:agenda}

\subsection{What the Evidence Establishes---and What It Does Not}

That the SDLC \emph{has been} reshaped is supported broadly and consistently: mapping studies and surveys across data engineering for AI~\cite{heck2024data}, software engineering for AI-based systems~\cite{martinez2022software}, ML testing~\cite{zhang2022machine,riccio2020testing}, deployment~\cite{paleyes2022challenges}, and MLOps~\cite{kreuzberger2023mlops,testi2022mlops,symeonidis2022mlops} converge with industrial case studies~\cite{sculley2015hidden,amershi2019software,arpteg2018software,baylor2017tfx,shankar2024we,munappy2022data} on the same phase-level transformations. However, the evidence base is dominated by literature reviews, taxonomies, interview and survey studies, and single-organization experience reports; comparable experiments, participant-level statistics, and pooled effect sizes are largely absent. The \emph{direction} of the transformation is therefore well established while its \emph{magnitude}---the causal effect of DE--SE integration on efficiency, quality, reliability, and maintainability---remains insufficiently quantified. Human-centered constraints---collaboration, explainability, skills, ethics, legacy integration, and security---are repeatedly identified as central and unresolved~\cite{martinez2022software,sambasivan2021everyone,nahar2022collaboration,abrahao2025software}.

This paper inherits the limitations of its evidence base. As a conceptual synthesis it does not itself contribute new empirical data; the formal model of Section~\ref{sec:model} is descriptive rather than predictive; and the design propositions of Section~\ref{sec:framework}, while grounded in the cited literature, await evaluation against explicit quality criteria in real organizations.

\subsection{Research Agenda}

Five directions follow. \textbf{(1) Measurement:} develop and validate instruments for integrated-lifecycle maturity and data-plane delivery performance, enabling multi-organization variance studies of the Section~\ref{sec:researchmodel} model~\cite{john2021towards,breck2017ml,serban2020adoption,forsgren2018accelerate}. \textbf{(2) Controlled and longitudinal evidence:} move beyond experience reports to longitudinal case studies, quasi-experiments (before/after platform adoption with matched controls), and repository-mining studies of co-evolution among code, data, and model artifacts~\cite{barrak2021co}. \textbf{(3) LLM- and agent-specific lifecycle science:} establish rigorous foundations for eval design and validity, judge reliability, prompt versioning semantics, and the engineering of multi-agent artifacts, where practice runs far ahead of theory~\cite{zheng2023judging,chang2024survey,he2025llm}. \textbf{(4) Adaptive process design:} apply design-science research to construct and evaluate the framework of Section~\ref{sec:framework}---reference architectures, DSL and model-driven support~\cite{radler2024bridging}, and formally analyzable gate semantics~\cite{renggli2019continuous}---against explicit quality models. \textbf{(5) Sustainability and human factors:} quantify the economic, environmental, and human costs of continuous training and evaluation and design lifecycles that optimize them jointly with quality~\cite{tamburri2020sustainable,abrahao2025software}, alongside educational research on curricula producing engineers fluent across the integrated stack~\cite{kastner2020teaching}.

\section{Conclusion}
\label{sec:conclusion}

The integration of data engineering and software engineering practices does not merely add steps to the SDLC; it changes what the lifecycle \emph{is}. A linear, code-centric, deterministic process has become a continuous, multi-loop, probabilistic ecosystem in which code, data, and models are co-equal, co-versioned, co-tested, and co-deployed artifacts. Requirements become data-aware and executable as evaluation-led specifications; architecture becomes contract-first, organized around feature stores, registries, orchestrated pipelines, and lineage; development extends ``as code'' discipline and shift-left quality to the data plane; testing becomes a multi-layered stack of data, pipeline, model, behavioral, and evaluation gates wired into CI; deployment expands into CI/CD/CT/CM with progressive delivery and pipeline-aware rollback; maintenance becomes closed-loop drift control maintaining $c_t \in \accept(\spec_t)$; and governance becomes an engineering practice enforced by the same infrastructure that delivers the system. The literature establishes the reality and direction of this transformation beyond reasonable doubt, while leaving its measured magnitude, its optimal designs, and its human dimensions open---precisely the space in which the proposed research model and adaptive-lifecycle framework are positioned. In the integrated lifecycle, software engineering provides the rigor, scalability, and reliability, while data engineering ensures that the fuel (data) and the engine (models) are continuously optimized and aligned; systems built this way are more iterative, automated, observable, and governed than their predecessors---and also more complex and perpetually evolving, which is exactly why they need a lifecycle engineered to adapt.

\bibliographystyle{unsrt}
\bibliography{refs}

\end{document}